\documentclass[12pt]{iopart}
\usepackage{epsfig}

\begin{document}
\title[Automated Reconstruction of Particle Cascades in HEP Experiments]
{Automated Reconstruction of Particle Cascades in High Energy Physics Experiments}

\author{
O Actis, 
M Erdmann, 
A Henrichs\footnote{Present address: 
II. Physikalisches Institut, Universit\"at G\"ottingen, Friedrich-Hund-Platz 1, D-37077 G\"ottingen}, 
A Hinzmann, 
M Kirsch, \\
G M\"uller, 
J Steggemann
}

\address{III. Physikalisches Institut A, RWTH Aachen University, Otto-Blumenthal-Str., D-52056 Aachen}
\ead{erdmann@physik.rwth-aachen.de}

\begin{abstract}
We present a procedure for reconstructing particle cascades from event data measured 
in a high energy physics experiment. 
For evaluating the hypothesis of a specific physics process causing the observed data, all possible reconstruction versions of the scattering process are constructed from the final state objects.
We describe the procedure as well as examples of physics processes of different complexity 
studied at hadron-hadron colliders. 
We estimate the performance by 20~$\mu$s per reconstructed decay vertex, 
and 0.6~kByte per reconstructed particle in the decay trees.
\end{abstract}



\submitto{Computational Science \& Discovery}

\maketitle

\section{Introduction}

When searching for heavy particles in high energy physics experiments,
one often needs to reconstruct the whole decay cascade of these particles.
This can be helpful, e.g., in order to prove the existence of a particle 
through its invariant mass distribution.

Since only the final state particles are measured in a detector, physicists
encounter several difficulties in the reconstruction of these cascades.
\begin{enumerate}
\item
Depending on the depth of the particle decay cascade, and the number of
final state particles, the number of possible reconstruction versions may
become rather large.
\item
If several decay channels for the heavy particle are analyzed, all
different cascades need to be reconstructed.
\item
The correctness of each reconstruction version of the cascade is to be 
evaluated in order to suppress false reconstruction versions.
\end{enumerate}
Regarding the first two items, source code for reconstructing these decay cascades 
in a data analysis is usually programmed manually for every physics process under 
consideration.
The third item requires specific knowledge of a physicist, taking all possible measures
to identify the correct reconstruction versions, and to use this information 
for the physics interpretation of the data.

In this paper we present a procedure for automated reconstruction of particle 
cascades.
Such a procedure accelerates the design phase of a physics analysis, 
as it avoids the above mentioned programming of every physics process individually.
It allows physicists to explore different physics interpretations of 
the data on a short time scale.
The challenges for the realization of such an automated procedure are
\begin{itemize}
\item
Flexibility in the physics process selected to analyze the data,
\item
Steering mechanism for reconstructing the required cascade,
\item
Management of multiple reconstruction versions of the cascades,
\item
Management of the large number of mother-daughter relations between the particles 
within the decay cascade, 
\item
Treatment of invisible particles, and
\item
Performance with respect to CPU time and memory consumption.
\end{itemize}

For our implementation we rely on the C++ toolkit Physics eXtension Library 
(PXL~\footnote{\it http://sourceforge.net/projects/pxl}) \cite{Steggemann}
which is a successor project of the Physics Analysis eXpert (PAX) package 
\cite{Erdmann:2002wn}-\cite{Kappler:2006uq}.
These packages have been designed to support physicists in physics analyses of any complexity.
Their main characteristics are a general event container enabling
users to store all information required in a physics data analysis including multiple
reconstruction versions of a single event, and management of relations, 
such as mother-daughter relations.

This publication is organized as follows. We first describe the 
technology with the steering part, the algorithm itself, and its output. 
Then we give several examples of physics processes, and present performance measures for each process.

\section{Reconstruction Technique}
\subsection{Steering Interface}

To steer the algorithm, a template of the cascade to be reconstructed is required. As the steering interface we use the event container provided in the PXL toolkit. The container is designed to hold all information needed in the analysis of a high energy physics collision event, e.g. particles, or data required for specific physics analyses. Particles are defined within the PXL software as well. Each particle carries a four-momentum, name, analysis-specific data, et cetera. Relations, e.g. between mother particles and daughter particles, can be established as well.

The steering event container is to be provided by a physicist. The event container is used to hold all particles which are expected in the cascade of the physics process, as well as their mother-daughter relations. This information is used by the algorithm to reconstruct the requested physics process from the objects measured in a detector.

\subsection{Event Data Interface}

To input the relevant data of a particle collision event into the algorithm, we also use the event container of the PXL software as interface. A physicist provides the corresponding event container which then holds the required reconstructed objects, e.g. muons, electrons, photons, or jets, in the form of PXL particles. In addition, the event container and its particles hold all other information needed for the physics analysis, such as missing transverse momentum, trigger information, bottom quark identification, et cetera.

\subsection{Reconstruction of the Particle Cascade}

The algorithm to reconstruct the particle cascade is based on an iterative procedure. 
In the first iteration, the steering event container is searched for two particles 
that do not decay further and have a common mother. 
In the event data, two reconstructed objects corresponding to the daughter particles 
are then taken to reconstruct a mother particle, observing four-momentum conservation.
Both the newly built mother and the mother-daughter relations are added to the 
reconstructed event data.

If more reconstructed objects of the correct type are found than required in the particle decay under consideration, multiple versions of the reconstructed event container are built. This multiplication procedure continues until all possible combinations of reconstructed objects are created to form the mother particle. 

The newly reconstructed mother particles are taken as being completed, and are then treated as non-decaying particles. Hereafter the procedure to find two particles that do not decay and have a common mother is continued until no further particle decay needs to be reconstructed.

In our C++ implementation of the algorithm as described in this paper, up to one final state particle in the steering template can be considered as having escaped detection, e.g. a neutrino. Then, the kinematics of the mother particle is calculated using the missing transverse momentum components of the event, and the mother particle mass as given in the steering event container. Depending on the solution of a quadratic equation based on a mass constraint, at most two versions of the cascade are determined. In case of an imaginary solution of the equation, only the real part is taken into account. 

\subsection{Output}

On output, the algorithm delivers all possible reconstruction versions using the PXL event container as the interface. 
Each container holds the originally reconstructed objects of the detector together 
with all reconstructed particles of the cascade and their corresponding mother-daughter relations.

\section{Performance}

\begin{figure}
\setlength{\unitlength}{1cm}
\begin{picture}(16.0,10.0)
\put(1.0,0.0){\epsfig{file=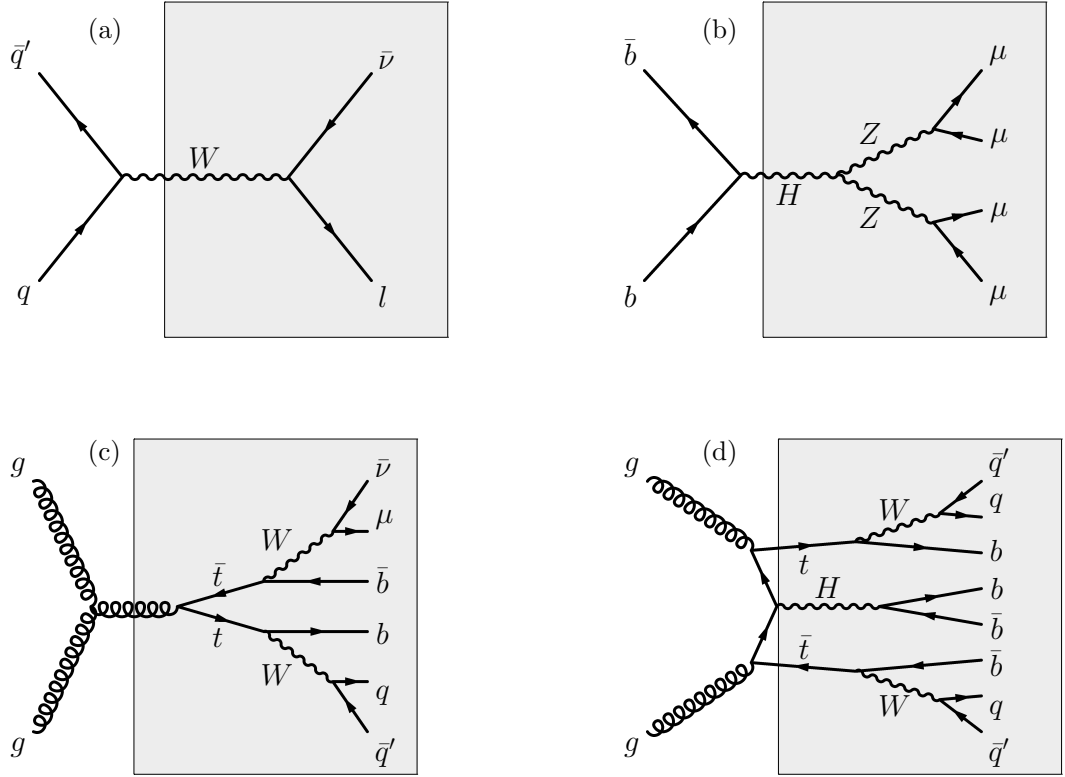,width=14.0cm}}
\end{picture}

\caption{Example Feynman diagrams of 
(a) $W\ $boson production with leptonic decay, 
(b) production of a Higgs boson decaying into four muons, 
(c) top-antitop quark pair production with decay into a muon, neutrino, and 4 jets, 
(d) top associated Higgs boson production with hadronic decays. 
The shaded areas indicate the particle cascades to be reconstructed.}
\label{fig:feyn}
\end{figure}

To check the performance of the algorithm in our implementation, we use different physics processes to measure the time and memory consumption when reconstructing all possible versions of the cascade decay. For a rather complete characterization of the algorithm, we use examples of different complexity (see figure~\ref{fig:feyn}).

In our test procedure we use events produced with the MadEvent generator 
V4.1.33 \cite{Maltoni:2002qb} applying a setup suitable for the 
Large Hadron Collider (LHC).
They contain the corresponding example physics process with the true particle 
cascade known. 
We then construct pseudo-data by stripping off the cascade information, 
and by providing only the final state particles in an event container. 
Since we use the particles of the PXL package as the interface, 
the generated particles can in principle be replaced by leptons, jets, and other
objects measured in a detector.
The event container with the final state particles is then passed to the algorithm 
as the reconstructed event data.

\begin{figure}
\setlength{\unitlength}{1cm}
\begin{picture}(16.0,11.5)
\put(0.0,11.2){\epsfig{file=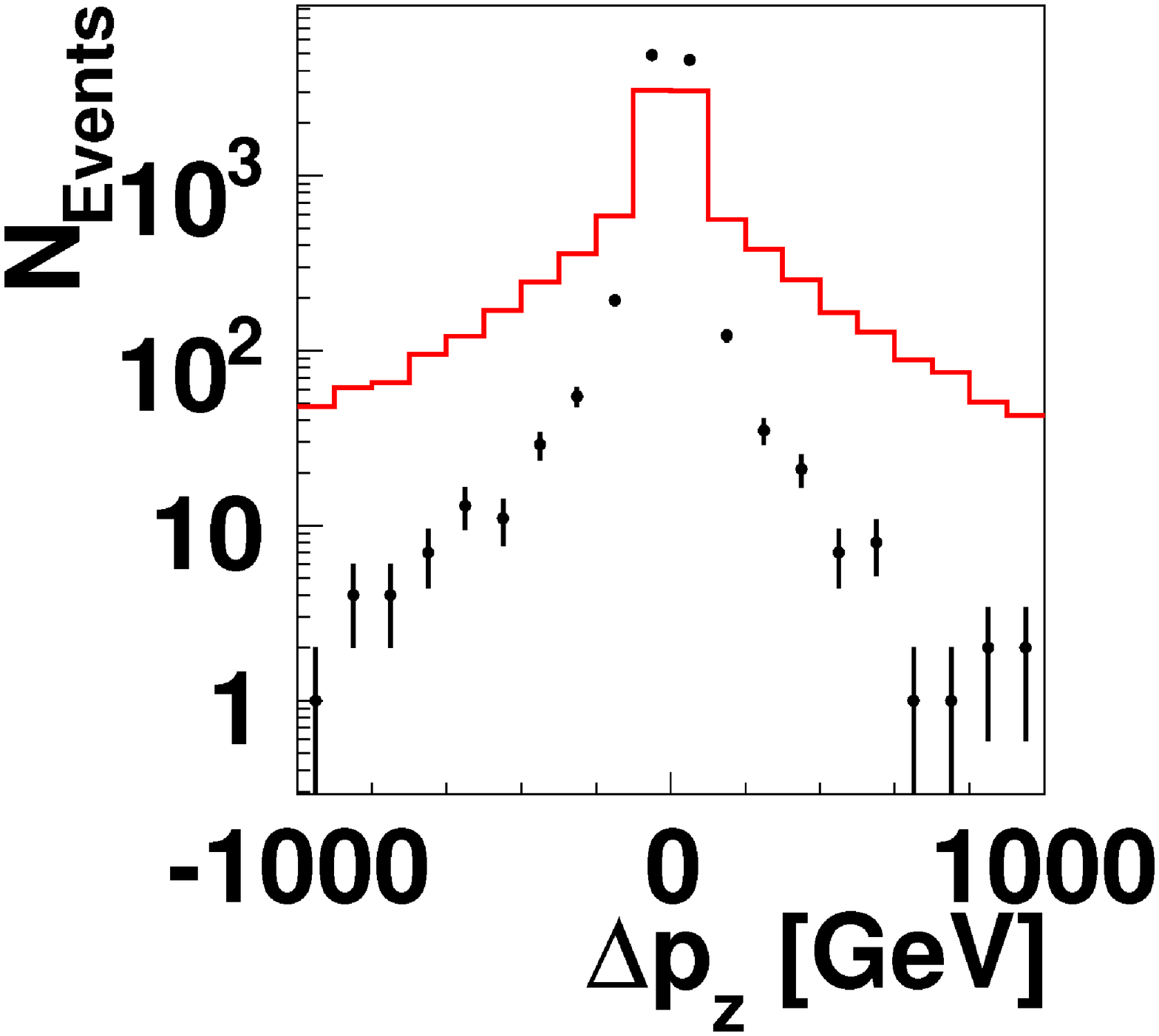,width=5.5cm,angle=-90}}
\put(7.5,11.2){\epsfig{file=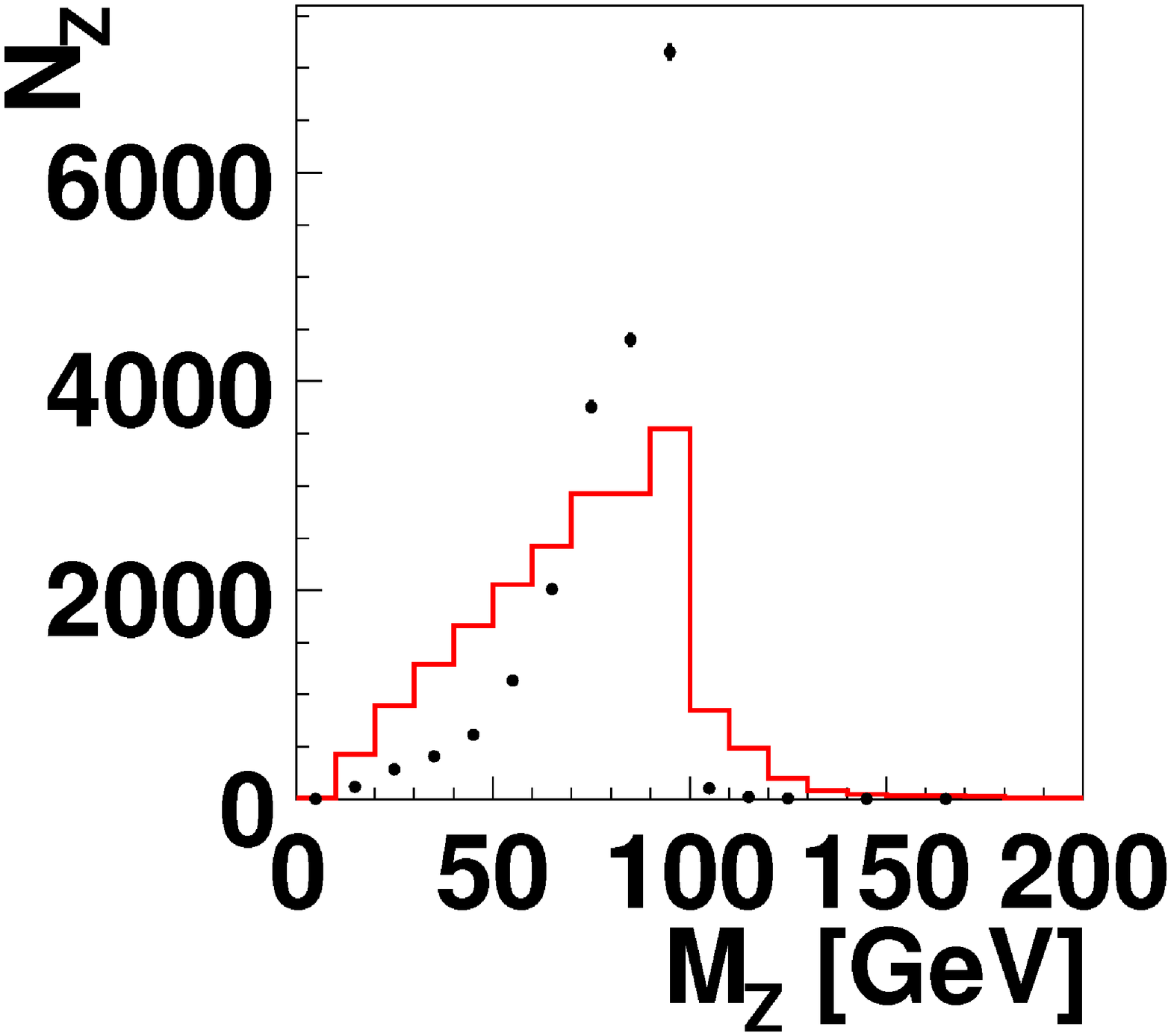,width=5.5cm,angle=-90}}
\put(0.0,5.2){\epsfig{file=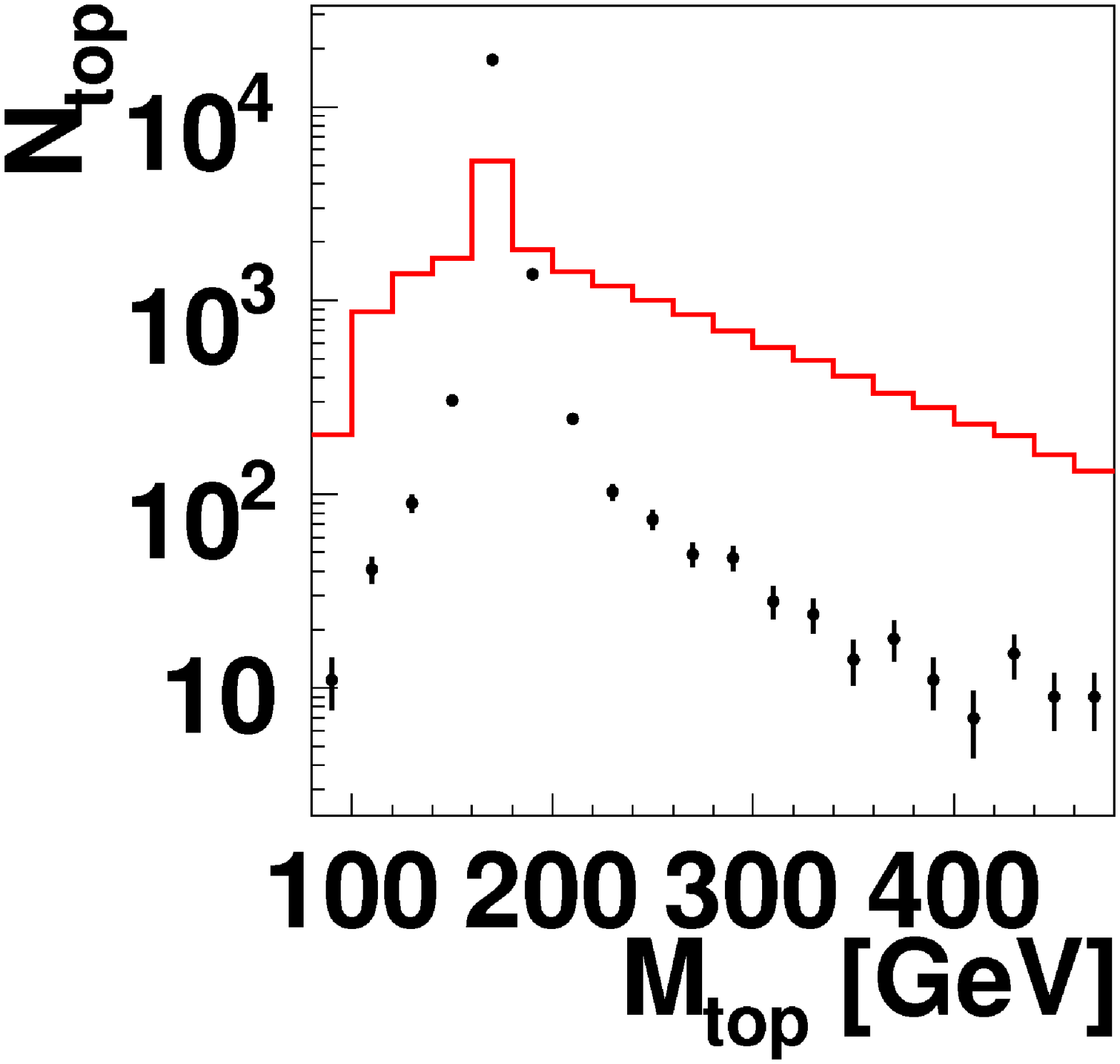,width=5.5cm,angle=-90}}
\put(7.5,5.2){\epsfig{file=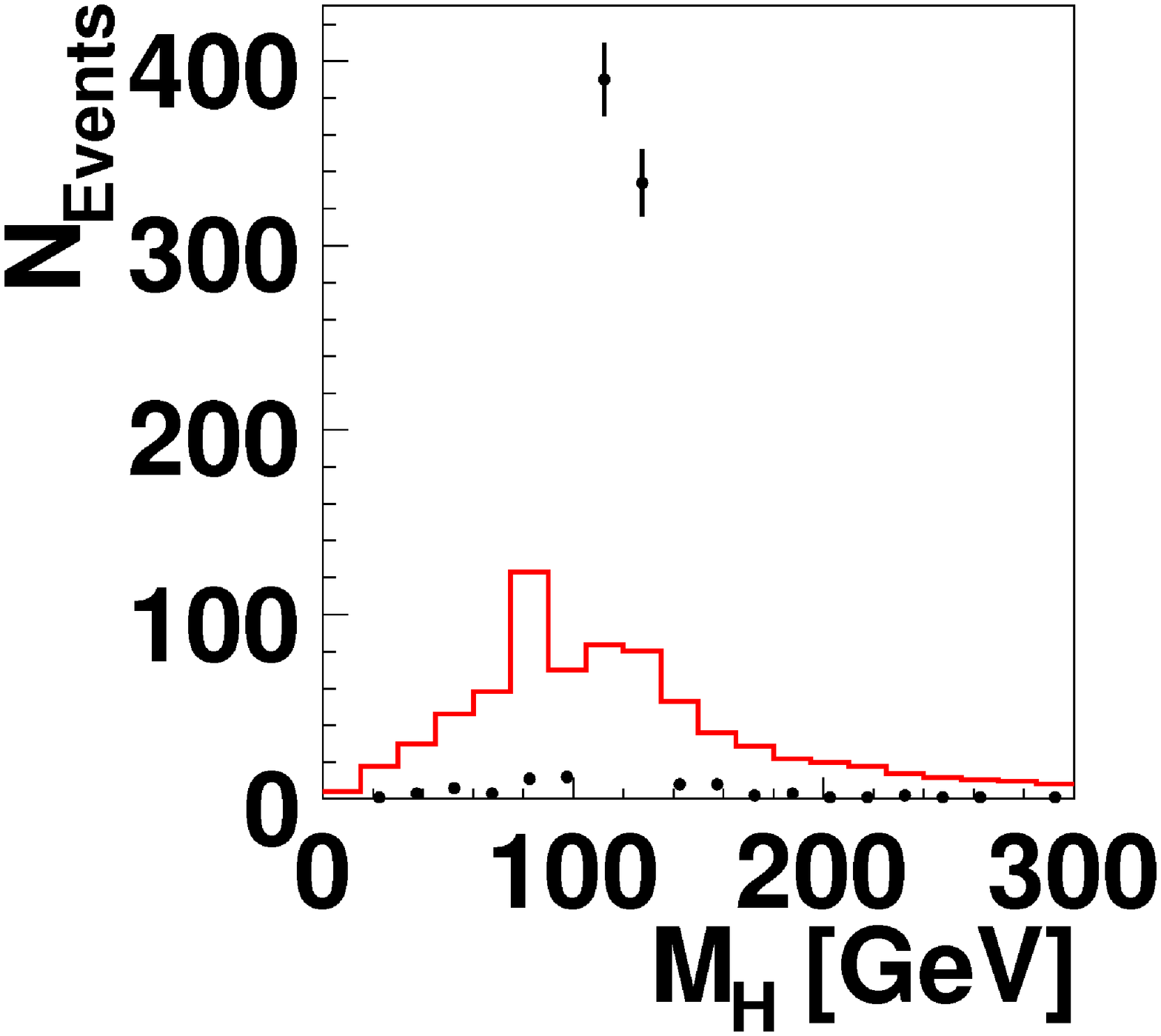,width=5.5cm,angle=-90}}
\put( 1.9,10.5){(a)}
\put( 9.4,10.5){(b)}
\put( 1.9,4.5){(c)}
\put( 9.4,4.5){(d)}
\end{picture}
\caption{Correctly reconstructed event (symbols), and all reconstructed decay cascades, weighted by the inverse of the number of reconstructed event versions (histograms). 
(a) Difference between reconstructed and generated longitudinal momentum of the neutrino in $W\ $boson processes, 
(b) mass of the $Z\ $bosons for processes with a Higgs boson decaying into two $Z\ $bosons, 
(c) mass of the top quarks in top pair production processes, and 
(d) Higgs boson mass in top associated Higgs production.}
\label{fig:reco}
\end{figure}

The histograms shown in figure~\ref{fig:reco} contain information from all reconstructed cascades, weighted by the inverse of the number of reconstructed event configurations. The symbols show the correctly reconstructed cascade obtained from a comparison with the true cascade.

In table \ref{tab_perf}, the measured performance values are shown for 
a computer performing at approximately 0.4~kSi2000.
The time values give the average time per event which the computer spent on the algorithm alone, measured over at least 1000 events. The memory size provides an estimate of the maximum allocated memory of the algorithm using the difference between the memory allocation with and without our algorithm.

Estimates for the average time and memory allocation are 20~$\mu$s per vertex, and 0.6~kByte per reconstructed particle in the decay trees, respectively.
These values have been validated from extended tests where additional final state particles were added to the top quark processes such that the number of reconstruction versions per event increased.

\begin{table}
\caption{\label{tab_perf}Performance measures of the reconstruction algorithm for particle cascades (see figure~\ref{fig:reco}). The first column indicates the physics process. Columns 2-4 show the number of particles, vertices, and possible reconstruction versions corresponding to the steering template. The last two columns show the average time per data event and the allocated memory, respectively.}
\begin{indented}
\item[]\begin{tabular}{@{}lrrrrr}
\br
Physics	 			& number of 	& number of	& number of	& Time/event & Mem. alloc.\\
process		    		& particles     & vertices	& rec. versions	&  [ms]      &  [MByte]   \\
\mr
$W\rightarrow l\nu$     	&   3		&  1		&    2 		&  0.06      &  $<$ 1    \\
$H\rightarrow 4\mu$		&   7		&  3		&    3 		&  0.4       &  $<$ 1	 \\
$t\bar{t}\rightarrow \mu\nu4j$  &  11		&  5		&   24 		&  2.3       &  $<$ 1	 \\
$Ht\bar{t}\rightarrow 8 j$	&  13		&  5		& 5040 		&  411       &  36	 \\
\br
\end{tabular}
\end{indented}
\end{table}

\subsection{$W\ $Boson Decay into Leptons}

$W\ $boson production is considered as a candidate process to measure luminosity at the LHC. An example Feynman diagram is shown in figure~\ref{fig:feyn}(a). For the steering template of the algorithm, we use the $W\ $boson, the charged lepton and the neutrino, as indicated by the shaded area in figure~\ref{fig:feyn}(a).

In this process, typically two solutions of the kinematics of the $W\ $boson are calculated using the missing transverse momentum components and the $W\ $mass as a constraint. The resulting distribution of the difference between reconstructed and generated longitudinal momentum of the neutrino is shown in figure~\ref{fig:reco}(a). The performance of the algorithm is 0.06~ms per event with a memory allocation of less than 1~MByte.

\subsection{Higgs Boson Decay into Four Muons}

One of the possible channels in searches for the Higgs boson predicted within the Standard Model is its decay into two $Z$ bosons, with each $Z$ decaying into two muons. An example Feynman diagram of the process is shown in figure~\ref{fig:feyn}(b). The shaded area indicates the particles used for the steering template. Disregarding the muon charges, the cascade can be reconstructed in three different versions, each giving different reconstructed $Z$ bosons in the intermediate state. Here the algorithm needs 0.4~ms per event, with less than 1~MByte memory allocated. Figure~\ref{fig:reco}(b) shows the resulting mass distribution of the $Z$ bosons.

\subsection{Top Pair Production in the Lepton Plus Jets Channel}

Owing to the large top quark mass, top quark production will be one of the most interesting Standard Model processes at the LHC. An example Feynman diagram for a top pair production process with the so-called lepton plus jets decay channel is shown in figure~\ref{fig:feyn}(c), with the part used for the steering template indicated by the shaded area. Like in $W\ $boson production, there is a two-fold ambiguity for the reconstruction of the neutrino longitudinal momentum. Taking the quark-jet associations without identification of bottom quarks into account, there are in total 24 possible reconstruction versions. The time per event is 2.3~ms with, again, less than 1~MByte of allocated memory. The resulting top mass distribution is shown in figure~\ref{fig:reco}(c).

\subsection{Top Quark Associated Higgs Boson Production}

Top quark associated Higgs boson production will be studied in the context of understanding the Higgs boson coupling strengths at the LHC. If all decay products of the top quarks and the Higgs boson are quarks, there are 5040 possibilities to reconstruct the event. Here we ignored identification of bottom quarks in the decay chain to test a rather extreme situation. Figure~\ref{fig:feyn}(d) shows an example Feynman diagram, the shaded area indicates the part used for the steering template. Each event takes on average 411~ms to process, and a memory of 36~MByte is allocated. The Higgs mass distribution is displayed in figure~\ref{fig:reco}(d).

\section{Summary}

In this work, we explored an automated way of reconstructing particle decay cascades. We implemented a C++ version of a procedure which enables reconstruction of all possible reconstruction versions of a decay chain from the final state objects measured in a detector. The algorithm performs according to a single template that represents the requested decay chain and is provided by the user. We validated the algorithm using several physics processes of different complexity, and verified that the performance values with respect to time and memory consumption are within the scope of a typical physics analysis in high energy physics experiments.

\ack

We are very grateful for financial support of the Ministerium f\"{u}r Innovation, Wissenschaft, Forschung und Technologie des Landes
Nordrhein-Westfalen, the Bundesministerium f\"{u}r Bildung und Forschung (BMBF), and the Deutsche Forschungsgemeinschaft (DFG).


\section*{References}
\begin{thebibliography}{10}

\bibitem{Steggemann}
Actis~O, Erdmann~M, Fischer~R, Kirsch~M, Klimkovich~T, M\"uller~G, Plum~M, and Steggemann~J
2008
Visual physics analysis VISPA - concepts and first applications
{\it To appear in Proc. Int. Conf. on High Energy Physics (ICHEP08) (Philadelphia, USA)}

\bibitem{Erdmann:2002wn}
Erdmann~M, Hirschb\"{u}hl~D, Kemp~Y, Schemitz~P, and Walter~T 
2002 
Physics analysis expert
{\it Proc. 14th Topical Conf. on Hadron Collider Physics (HCP 2002) (Karlsruhe, Germany)} (Springer Verlag, Berlin, Germany)

\bibitem{Erdmann:2003ke}
Erdmann~M, Hirschb\"{u}hl~D, Jung~C, Kappler~S, Kemp~Y, and Kirsch~M 
2003
Physics analysis expert PAX: first applications
{\it Proc. Int. Conf. for Computing in High-Energy and Nuclear Physics (CHEP 03) (La Jolla, California)}
({\it Preprint} physics/0306085)

\bibitem{chep04}
Erdmann~M {\it et al} 
2004 
New Applications of PAX in physics analyses at hadron colliders
{\it Proc. Int. Conf. on Computing in High Energy and Nuclear Physics (CHEP04) (Interlaken, Switzerland)}

\bibitem{Kappler:2005tf}
Kappler~S, Erdmann~M, Felzmann~U, Hirschb\"uhl~D, Kirsch~M, Quast~G, Schmidt~A, and Weng~J 
2006
The PAX toolkit and its applications at Tevatron and LHC
{\it IEEE Trans.\ Nucl.\ Sci.}  {\bf 53} 506 
({\it Preprint} physics/0512232)

\bibitem{Kappler:2006uq}
Kappler~S, Erdmann~M, Kirsch~M, M\"uller~G, Weng~J, Flossdorf~A, Felzmann~U, Quast~G, Saout~C and Schmidt~A
2006 
Concepts, developments and advanced applications of the PAX toolkit
{\it Proc. Int. Conf. on Computing in High Energy and Nuclear Physics (CHEP06) (Mumbai, India)}
({\it Preprint} physics/0605063)

\bibitem{Maltoni:2002qb}
Maltoni~F and Stelzer~T 
2003 
MadEvent: Automatic event generation with MadGraph 
{\it JHEP} {\bf 0302} 027 ({\it Preprint} hep-ph/0208156)

\end{thebibliography}
\end{document}